\documentclass[aps,prl,twocolumn,showpacs,amsmath,amssymb,groupedaddress]{revtex4-1}
\usepackage{graphicx}
\begin{document}
\title{Tuning the Tricritical Point with Spin-orbit Coupling in Polarized Fermionic Condensates}
\author{Renyuan Liao}\email{rliao08@gmail.com}
\author{Yu Yi-Xiang}\email{yyx@iphy.ac.cn}
\author{Wu-Ming Liu}

\affiliation{National Laboratory for Condensed Matter Physics, Institute of Physics, Chinese Academy of Sciences, Beijing 100190, China}

\date{\today}

\begin{abstract}
 We investigate a two-component atomic Fermi gas with population imbalance in the presence of Rashba-type spin-orbit coupling (SOC). As a competition between SOC and population imbalance, the finite-temperature phase diagram reveals a large varieties of new features, including the expanding of the superfluid state regime and the shrinking of both the phase separation and the normal regimes. For sufficiently strong SOC, the phase separation region disappears, giving way to the superfluid state. We find that the tricritical point moves toward regime of low temperature, high magnetic field, and high polarization as the SOC increases.
\end{abstract}
\maketitle
Recent experimental realization of synthetic gauge field~\cite{LIN09} and spin-orbit coupling~\cite{LIN11} opens a new arena to explore quantum many-body systems of ultracold atoms. The engineered SOC (with equal Rashba and Dresslhaus strength) in a neutral atomic  BEC was accomplished by dressing two atomic spin states with a pair of lasers. The interaction between a quantum particles' spin and its momentum is crucial for spin Hall effects~\cite{XIA10} and topological insulators~\cite{HAS10}, which has captured great attention in the condensed matter community. The engineered SOC, equally applicable for bosons and fermions, allows for the realization of topological insulators and topologically nontrivial states~\cite{SAT09,SAU11} in fermionic neutral atom systems, engendering broad interests in physical community.

In anticipation of immediate experimental relevance involving SOC in fermionic atoms such as $^6Li$ and $^{40}K$\cite{SAU11}, intense theoretical attention has been paid to the physics of BEC-BCS crossover~\cite{VYA11,CHU11,ZHA11,HUI11} and polarized Fermi gases~\cite{ISK11,HAN11,GUO11} in the presence of SOC . The SOC has been predicted to lead to various new phenomena. In particular, for the two-body problem, it gives rise to a two-body bound state even on the BCS side ($a_s<0$) of a resonance~\cite{JAY11}. For the many-body physics at mean-field level, it enhances BCS pairing via the increased density of states at low energy, and leads to anisotropic superfluids through mixing the spin singlet and triplet components~\cite{GOR01,HUI11}.

Polarized fermionic condensates have been the focus of both theoretical and experimental research over the past years~\cite{GIO08}.
One of the key questions to ask is that how SOC reshape our understanding of these exciting systems. So far, most of the theoretical studies~\cite{CHU11,ZHA11,HUI11,ISK11,HAN11,GUO11} have focused mainly on zero temperature, leaving the physics at finite temperature which is experimentally relevant largely intact. In this paper, we are trying to addressing this question by conducting the following studies: Firstly, we map out the $\textit{finite-temperature}$ phase diagram at BCS side where mean-field theory gives quantitatively reasonable results. By determining the behavior of the tricritical point
as a function of SOC strength, we can completely characterize the topology of the phase diagram without recourse to
an extensive numerical treatment~\cite{PAR07,PAR072}; Secondly, we examine the physics at unitary, which is experimentally relevant and theoretically interesting. Specifically , we consider the effect of SOC on the ``spin susceptibility" and the critical temperature. Finally, we investigate the fate of breached pairing states~\cite{LIU03} under SOC through the correlation functions.


We consider a homogeneous two species polarized Fermi gases  interacting via an attractive contact potential with an isotropic in-plane Rashba spin-orbit coupling, described by the following Hamiltonian,
\begin{eqnarray}
 H&=&\int d^3\mathbf{r}\sum_{\sigma=\uparrow,\downarrow}\psi_\sigma^\dagger(\mathbf{r})(\frac{{\hat{\mathbf{P}}}^2}{2m}-\mu_\sigma)\psi_\sigma(\mathbf{r})\nonumber\\
 & &-g\int d^3\mathbf{r}\psi_\uparrow^\dagger(\mathbf{r})\psi_\downarrow^\dagger(\mathbf{r})\psi_\downarrow(\mathbf{r})\psi_\uparrow(\mathbf{r})+H_{SO}.
\end{eqnarray}
Here, $H_{SO}=\lambda\sum_\mathbf{k}k_\perp\left[ e^{-i\varphi_\mathbf{k}}c_{\mathbf{k}\uparrow}c_{\mathbf{k}\downarrow}+h.c.\right]$, with the transverse momentum $\bold{k_\perp}=(k_x,k_y)$ and $\varphi_\mathbf{k}=Arg(k_x+ik_y)$. The strength of spin-orbit coupling $\lambda$ can be tuned by atom-laser interaction~\cite{LIN11}. We define the chemical potential $\mu$ and ``Zeeman" field $h$ such that $\mu_\uparrow=\mu+h$ and $\mu_\downarrow=\mu-h$. The spin imbalance between the two species is denoted by the polarization $P=(n_\uparrow-n_\downarrow)/(n_\uparrow+n_\downarrow)$. We consider pairing between different hyperfine species of the same atom, so we restrict ourself to a single mass m. The interaction strength $g$ is expressed in terms of the s-wave scattering length $a_s$ using the prescription:
$\frac{m}{4\pi a_s}=-\frac{1}{g}+\frac{1}{V}\sum_\mathbf{k}\frac{1}{2\epsilon_\mathbf{k}}$,
where $V$ is the volume and $\epsilon_\mathbf{k}=\mathbf{k}^2/2m$ (for convenience, we set $\hbar=k_B=1$). We also define the Fermi momentum using $k_F=(3\pi^2n)^{1/3}$, with total density $n=n_\uparrow+n_\downarrow$,  so that the Fermi velocity is $v_F=k_F/m$. Throughout our calculation, we will keep $n$ fixed.

Within the framework of imaginary-time field integral, the partition function of the system is $Z=\int d[\overline{\psi},\psi]\exp{(-S[\overline{\psi},\psi])}$ with the action $ S[\overline{\psi},\psi]=\int d\tau\left[\sum_\sigma\overline{\psi}_\sigma\partial_\tau\psi_\sigma+H(\overline{\psi},\psi)\right]$.
Introducing a bosonic field $\Delta(\mathbf{r},\tau)$, which is believed to encapsulate the relevant low-energy degrees of freedom, we perform a Hubbard-Stratonovich transformation, then the action becomes

$S=\int d\tau[\sum_\mathbf{k\sigma}(\epsilon_\mathbf{k}-\mu_\sigma)\overline{\psi}_{\mathbf{k}\sigma}\psi_{\mathbf{k}\sigma}+H_{SO}]
  +\int d\tau d^3\mathbf{r}\left(\frac{|\Delta^2|}{g}-\Delta\overline{\psi}_\uparrow\overline{\psi}_\downarrow-\overline{\Delta}\psi_\downarrow\psi_\uparrow\right)$.
To bring the action in a compact form, we define a four-dimensional vector $\overline{\Psi}_\mathbf{k}=\left(\overline{\psi}_{\mathbf{k}\uparrow}\overline{\psi}_{\mathbf{k}\downarrow}\psi_{-\mathbf{k}\uparrow}\psi_{-\mathbf{k}\downarrow}\right)$. Then the action can be casted as $ S=\int d\tau \sum_{\mathbf{k}}\left[\frac{1}{2}\overline{\Psi}_\mathbf{k}(-\mathcal{G}^{-1})\Psi_\mathbf{k}+\xi_\mathbf{k}\right]$,
with the inversed Green's function defined as
\begin{eqnarray*}
   \mathcal{G}^{-1}=\begin{pmatrix}
                      -\partial_\tau-\xi_{\mathbf{k}\uparrow}& -\lambda k_\perp e^{-i\varphi_\mathbf{k}}&0 &\Delta\\
                      -\lambda k_\perp e^{i\varphi_\mathbf{k}} & -\partial_\tau-\xi_{\mathbf{k}\downarrow} & -\Delta & 0\\
                      0 & -\overline{\Delta} & -\partial_\tau+\xi_{\mathbf{k}\uparrow} & -\lambda k_\perp e^{i\varphi_\mathbf{k}}\\
                      \overline{\Delta}& 0& -\lambda k_\perp e^{-i\varphi_\mathbf{k}} & -\partial_\tau+\xi_{\mathbf{k}\downarrow}
                    \end{pmatrix},
\end{eqnarray*}
with $\xi_{\mathbf{k}\uparrow}=\epsilon_\mathbf{k}-\mu+h$, $\xi_{\mathbf{k}\downarrow}=\epsilon_\mathbf{k}-\mu-h$ and $\xi_\mathbf{k}=\epsilon_\mathbf{k}-\mu$ . Integrating out the fermionic degrees of freedom, we obtain the effective action
\begin{eqnarray}
 S_{eff}=\int d\tau d^3\mathbf{r}\frac{|\Delta|^2}{g}-\frac{1}{2}Tr\ln{(-\mathcal{G}^{-1})}+\beta\sum_\mathbf{k}\xi_\mathbf{k}.
\end{eqnarray}
Setting $\Delta=\Delta_0+\delta\Delta$, and $\mathcal{G}_0^{-1}=\mathcal{G}^{-1}|_{\Delta=\Delta_0}$, we can write $\mathcal{G}^{-1}=\mathcal{G}_0^{-1}+\Sigma$. Expanding the effective action to the second order in the fluctuation $\Sigma$, we approximate the effective action to be $S_{eff}\approx S_0+S_g$, with
$S_0=\beta V\frac{|\Delta_0|^2}{g}+\sum_{\mathbf{k}s=\pm}\left[\frac{\beta}{2}(\xi_\mathbf{k}-E_{\mathbf{k}s})-\ln{(1+e^{-\beta E_{\mathbf{k}s}})}\right]$,
$S_g=\beta V\sum_q\frac{\delta\overline{\Delta}(-q)\delta\Delta(q)}{g}
+\frac{1}{4}Tr\left[\mathcal{G}_0(k)\Sigma(-q)\mathcal{G}_0(k-q)\Sigma(q)\right]
\equiv\beta V\sum_q\Gamma^{-1}(q)\overline{\Delta}(-q)\delta\Delta(q)$,
where $k=(\mathbf{k},iw_n)$, $q=(\mathbf{q},i\nu_n)$, and the quasiparticle excitation spectrum $E_\mathbf{k\pm}$ determined by  $E_{\mathbf{k}\pm}^2=\xi_\mathbf{k}^2+\Delta_0^2+h^2+\lambda^2k_\perp^2\pm\sqrt{\xi_\mathbf{k}^2h^2+\xi_\mathbf{k}^2\lambda^2k_\perp^2+h^2\Delta_0^2}$ . While $E_{\mathbf{k}+}$ is always gapped, $E_{\mathbf{k}-}$ accommodates gapless excitations distributed symmetrically along $k_z=0$ axis at $k_\perp=0$: (1) for $\mu\geqslant\sqrt{h^2-\Delta_0^2}$, it has four gapless excitation points at $k_z=\pm\sqrt{\mu\pm\sqrt{h^2-\Delta_0^2}}$; (2) for $\mu<\sqrt{h^2-\Delta_0^2}$, it only has two gapless excitation points at $k_z=\pm\sqrt{\mu-\sqrt{h^2-\Delta_0^2}}$.

\begin{figure}[t]
{\scalebox{0.33}{\includegraphics[clip,angle=0]{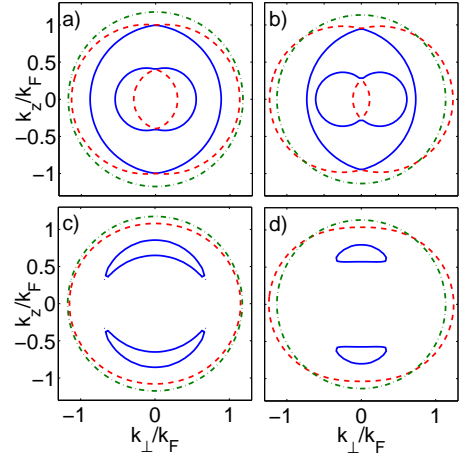}}}
\caption{(Color online) Isoenergy surface ($E_{\mathbf{k}\pm}=0.8E_F$) for the quasiparticle excitation spectrum at unitarity where $1/k_Fa=0$ at $T=0$:  (a) $h=0$, $\lambda=0.125v_F$; (b) $h=0$, $\lambda=0.25v_F$; (c) $h=0.1E_F$; $\lambda=0.125v_F$; (d) $h=0.1E_F$; $\lambda=0.25v_F$. The red dash line is plotted for $E_{\mathbf{k}+}$, the blue solid line is for $E_{\mathbf{k}-}$, and the green dash dotted circle is for a spherical fermi surface, plotted for comparison.
}
\label{Fig1}
\end{figure}

Peculiar properties of the excitation spectrum are illustrated in Fig.~\ref{Fig1}, where the isoenergy surface for $E_{\mathbf{k}\pm}=0.8E_F$ at unitary ($1/k_Fa=0$) is shown at zero temperature. The red dash curve is for $E_\mathbf{k-}$, the blue solid curve is for $E_\mathbf{k+}$, and the green dash dotted circle is for a spherical Fermi surface. The isoenergy surface is symmetric with respect to $k_z=0$, and possesses rotation symmetry along $z$-axis. For balanced superfluid ($h=0$), $E_{\mathbf{k}\pm}=\sqrt{\left(|\xi_\mathbf{k}|\pm\lambda k_\perp\right)^2+\Delta_0^2}$, as depicted in panel (a) and (b), with $\lambda=0.125v_F$ and $\lambda=0.25v_F$ respectively. The anisotropy of the isoenergy surface increases as one increases the strength of SOC $\lambda$. It is interesting to notice that there exists two branches of isoenergy for both  $E_\mathbf{k+}$ and $E_\mathbf{k-}$, due to the positiveness of chemical potential in this case. This will lead to the enhancement of BCS pairing through increasing density of states around Fermi surface. $E_{\mathbf{k}+}$ and $E_{\mathbf{k}-}$ merges at $k_\perp=0$ at which effects of SOC vanishes.  For $h=0.1E_F$, the isoenergy surface is shown in panel (c) and (d), with $\lambda=0.125v_F$ and $\lambda=0.25v_F$ respectively. Here we only have one branch for both $E_\mathbf{k+}$ and $E_\mathbf{k-}$. Interestingly, the curve for $E_\mathbf{k+}$ develops a lunar structure for low $\lambda$, as shown in panel (c), and increasing $\lambda$ makes it blunt.

\begin{figure}[t]
{\scalebox{0.33}{\includegraphics[clip,angle=0]{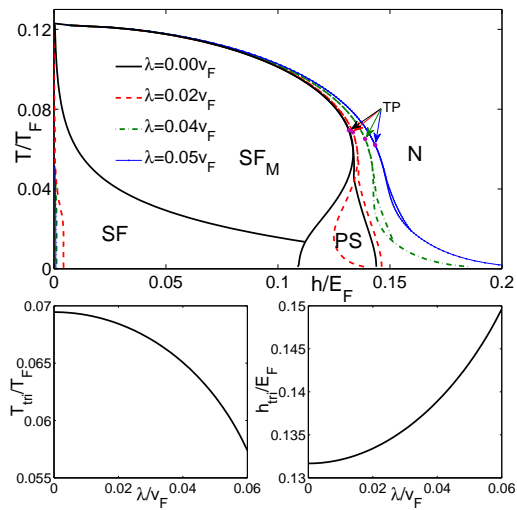}}}
\caption{(Color online) Upper panel: finite-temperature phase diagram as a function of $T$ and $h$ at $1/k_Fa_s=-1$ (BCS side). There are four different phases: the N state, the PS state, the SF state, and the magnetized superfluid (SF$_M$). Above the tricritical point, the transition line separating the broken-symmetry state (SF$_M$) and the symmetric state (N) is of second order.  Below the tricritical point (TP) it changes to the first order. Lower panel: the evolution of the tricritical point ($T_{tri}/T_F$, $h_{tri}/E_F$) as a function of SOC strength $\lambda$. }
\label{Fig2}
\end{figure}
 Phase diagram at finite temperature plays a key role in characterizing polarized fermionic condensates~\cite{PAR07,KET08}. For a fixed interaction strength $1/k_Fa$ and a fixed SOC strength $\lambda$, the phase diagram  could be determined by the plane spanned by the temperature $T=1/\beta$ and the Zeeman field $h$. At sufficient low polarization or chemical potential difference h we expect a finite-temperature phase transition at which the superfluid order parameter vanishes continuously. Conversely, at low temperature, superfluidity is destroyed in a first order fashion with increasing h. Across this phase transition at fixed h, the polarization jumps discontinuously. To determine the position of the phase boundaries, we must minimize the mean-field grand potential $\Omega_0=S_0/\beta$ with respect to the BCS order parameter $\Delta_0$. Such a mean-field analysis should provides a qualitative reasonable description at weak-coupling BCS regime. In the absence of SOC ($\lambda=0$), it is well known that there exists a finite-temperature tricritical point in the BCS limit, which is a natural consequence of having a first order transition from the superfluid phase (SF) to the normal phase (N) at $T=0$ and a second order transition at zero polarization. First investigated by Sarma~\cite{SAR63} in the context of superconductivity in the presence of a magnetic field $h$, the BCS tricritical is located at $(T_{crit}/\Delta_0,h_{crit}/\Delta_0)=(0.3188,0.6061)$~\cite{CAS03}, where $\Delta_0=8/e^2E_F\exp{(-\pi/2|k_Fa|)}$. The phase diagram spanned by $T$ and $h$ at $1/k_Fa=-1$ for various SOC strength is shown in the upper panel of Fig.~\ref{Fig2}. It consists of four different phases: the superfluid state with zero polarization (SF), the magnetized superfluid state (SF$_M$), the normal state (N), and the phase separation (PS) regime enclosed by the the first order line and the second order line. As the strength of SOC $\lambda$ increases, the area of phase separation region diminishes, and eventually disappears for sufficient large $\lambda$. With the increasing of SOC, the regime of SF diminishes very sharply, giving way to SF$_M$. The intersection of the second order line and the first order line gives the position of the tricritical point, denoted as TP in Fig.~\ref{Fig2}. The evolution of the tricritical point is shown in the down panel. As $\lambda$ increases, $T_{tri}$ decreases, while $h_{tri}$ increases.
 \begin{figure}[t]
{\scalebox{0.33}{\includegraphics[clip,angle=0]{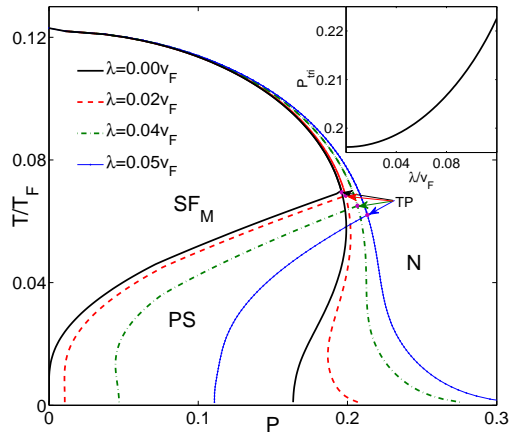}}}
\caption{(Color online) Finite-temperature phase diagram in the plane of $T$ and $P$ at $1/k_Fa_s=-1$. The inset shows the corresponding polarization $P_{tri}$ for the tricritical point as a function of SOC strength $\lambda$. The phase SF is along the line of $P=0$. The notation is the same as in Fig.~\ref{Fig2}.}
\label{Fig3}
\end{figure}

The finite temperature phase diagram for $1/k_Fa_s=-1$ spanned by $T$ and $P$ is shown in Fig.~\ref{Fig3}. The effect of SOC is dramatic at zero temperature: the system is unstable to phase separation at any polarization without SOC; however, as the SOC is turned on,  the system is a ``magnetized" superfluid, in which the superfluid component and the normal component coexist in an isotropic and homogeneous fashion. At finite temperature, by increasing the SOC strength $\lambda$, both the regions of normal and the phase separation diminishes, leaving the broadened regime of SF. The tricritical point moves towards high polarization and low temperature when $\lambda$ is increased, as seen in the inset.

The effect of SOC on the stability of the system manifests itself on the spin susceptibility of the system. We examine the stability of the superfluid phase at unitarity at zero temperature where mean-field theory should give qualitative reasonable arguments. As shown in Fig.~\ref{Fig3}, at $\lambda=0$, the system is unstable to phase separation as the slope of $\delta P/\delta h$ is always either zero or negative. There exists a critical polarization (here $P_c=0.68)$, above which the system reverts to the normal state. When the SOC is turned on, the superfluid state could support both low polarization and high polarization, in contrast to what we saw  Fig.~\ref{Fig2} at $1/k_Fa=-1$ where the superfluid state only support low polarization. At a sufficient large $\lambda$, the slope of the whole curve becomes positive, indicating that it is able to sustain any polarization.
\begin{figure}[t]
{\scalebox{0.38}{\includegraphics[clip,angle=0]{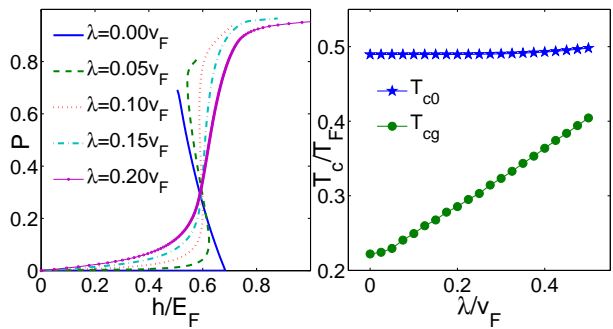}}}
\caption{(Color online) Left panel: the polarization $P=\frac{n_\uparrow-n_\downarrow}{n_\uparrow+n_\downarrow}$ as a function of magnetic field $h$ for various SOC strength $\lambda$ at zero temperature at unitarity. Right panel: the critical temperature for balanced superfluid at unitarity: $T_{c0}$ is calculated from mean-field theory and $T_{cg}$ is calculated by taking account of the Nozie$\acute{r}$es Schmitt-Rind correction.}
\label{Fig4}
\end{figure}

One of the most important questions to ask about the unitary superfluid is that how the critical temperature varies with SOC strength. At finite temperature, the contribution from non-condensed pairs to the density $n=\partial\Omega/\partial\mu$ becomes important. This contribution is necessary to approach the transition temperature of an ideal Bose gas in the molecular limit where $T_{BEC}=0.218T_F$ and can be included in the non-condensed phase ($\Delta_0=0$) through the gaussian contribution to the grand potential: $\Omega_g=(1/2\beta V)\sum_{\mathbf{q},i\nu_m}\ln{\Gamma^{-1}(\mathbf{q},i\nu_m)}$, where $\nu_m$ denotes the bosonic Matsubara frequencies. For numerical convenience, we treat the fluctuation by adopting the NSR scheme~\cite{NSR85}. The critical temperature for a balanced superfluid is shown in Fig.~\ref{Fig4} at unitarity.  The critical temperature calculated from mean-field theory $T_{c0}$, starts at a high value and increases slowly at small SOC. Taken account of the gaussian fluctuation, the critical temperature starts at about $0.224T_F$ and increases almost linearly with increasing SOC strength $\lambda$ for low $\lambda$, hinting at a possible way of realizing high-$T_c$ superfluids.

\begin{figure}[t]
{\scalebox{0.35}{\includegraphics[clip,angle=0]{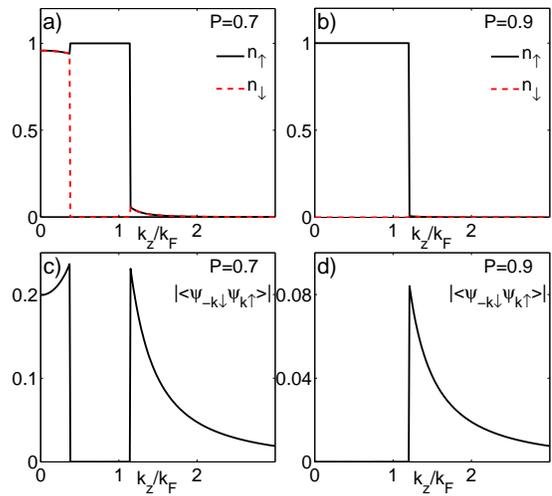}}}
\caption{(Color online) The momentum distribution $n_{\mathbf{k}\sigma}$ and correlation function $C_{\downarrow\uparrow}(\mathbf{k})=|<\psi_{-\mathbf{k}\downarrow}\psi_{\mathbf{k}\uparrow}>|$ at unitarity  at zero temperature with SO coupling strength $\lambda=0.2v_F$ for two typical polarization: $P=0.7$ (left panel); $P=0.9$ (right panel).}
\label{Fig5}
\end{figure}

Another interesting question concerning two-species spin imbalanced Fermi gases is how the picture of breached pairing state~\cite{LIU03} get modified. Breached pairing is characterized by a phase separation in momentum space between the excess of majority species $\uparrow$ and the minority species $\downarrow$ in the superfluid state. Signatures of phase separation are visible in the momentum distribution, $n_{\mathbf{k}\sigma}$, and correlation function $C_{\downarrow\uparrow}(\mathbf{k})=|<\psi_{-\mathbf{k}\downarrow}\psi_{\mathbf{k}\uparrow}>|$. Requiring the pairing amplitude to be zero, one finds the region for phase separation: (1) $k_\perp=0$; and (2) $|k_z|\in[0,\sqrt{\mu+\sqrt{h^2+\Delta_0^2}}]$ if $\mu<=\sqrt{h^2+\Delta_0^2}$ or $|k_z|\in[\sqrt{\mu-\sqrt{h^2+\Delta_0^2}},\sqrt{\mu+\sqrt{h^2+\Delta_0^2}}]$ if $\mu>\sqrt{h^2+\Delta_0^2}$.   Referring to Fig.~\ref{Fig5}, for $P=0.7$ shown in panel (a) and (c), there exists two typical momenta $k_{c1}$ and $k_{c2}$ between which the minority species $\downarrow$ is depleted and the majority species has full occupation, reminiscent of breached pairing state with two Fermi surfaces($BP_2$); while for $P=0.9$ shown in panel (b) and (d),  there exists a typical momentum $k_c$ below which the minority species is depleted and the majority species becomes fully occupied, reminiscent of $BP_1$. In both cases, the correlation function shows a ``hole" for momenta less than the Fermi momentum of the majority quasiparticles. The momentum distribution and the pairing amplitude bear consequences for experimental observation. The single-particle momentum distribution of trapped Fermi gases is routinely observed by time-of-flight measurements~\cite{REG05}.

To summarize, we have identified a series of new features arising from spin-orbit coupling. We hope that current work will add new excitement to the surging field of cold atom physics involving artificial gauge field and spin-orbit coupling.

 We are grateful to Han Pu, Qi Zhou and Fei Zhou for helpful discussions. This work was supported by NSFC under grants Nos. 10934010, 60978019, the NKBRSFC under grants Nos. 2009CB930701, 2010CB922904, 2011CB921502, 2012CB821300, and NSFC-RGC under grants Nos. 11061160490 and 1386-N-HKU748/10.
\bibliographystyle{apsrev4-1}
%

\end{document}